\documentclass[12pt,preprint]{aastex}
\usepackage[numberedappendix]{emulateapj5}  

\slugcomment{Accepted for the Astrophysical Journal}

\newcommand\be{\begin{equation}}
\newcommand\ee{\end{equation}}
\newcommand\bea{\begin{eqnarray}}
\newcommand\eea{\end{eqnarray}}
\newcommand\etal{{et al.}\thinspace}
\newcommand\lTr{{$ \langle T \rangle$}\thinspace}
\newcommand\lLr{{$ \langle L \rangle$}\thinspace}

\received{2004 January 9}
\begin{document}

\title{Entropy ``floor" and effervescent heating of intracluster gas}
\author{S. Roychowdhury\altaffilmark{1}, M. Ruszkowski\altaffilmark{2,4},
B. B. Nath\altaffilmark{1} and Mitchell C. Begelman\altaffilmark{2,3}}

\altaffiltext{1}{Raman Research Institute, Sadashivanagar, Bangalore
-560080, India; suparna,biman@rri.res.in}
\altaffiltext{2}{JILA, Campus Box 440,
University of Colorado at Boulder, CO 80309-0440; mr@quixote.colorado.edu, 
mitch@jila.colorado.edu}
\altaffiltext{3}{Department of Astrophysical and Planetary Sciences,
University of Colorado at Boulder}
\altaffiltext{4}{{\it Chandra} Fellow}

\begin{abstract}
Recent X-ray observations of clusters of galaxies have shown that the
entropy of the intracluster medium (ICM), even at radii as large as half
the virial radius, is higher than that expected from gravitational
processes alone. This is thought to be the result of nongravitational
processes influencing the physical state of the ICM. In this paper, we
investigate whether heating by central AGN can explain the distribution
of excess entropy as a function of radius. The AGN are assumed to inject
buoyant bubbles into the ICM, which heat the ambient medium by doing $pdV$
work as they rise and expand.  Several authors have suggested that this
``effervescent heating" mechanism could allow the central regions of
clusters to avoid the ``cooling catastrophe''. Here we study the effect of
effervescent heating at large radii. 
We find that the results are mainly sensitive to the total energy
injected into the cluster. Our calculations show that such a
heating mechanism is able to solve the entropy problem, provided that the
total energy injected by AGN is roughly proportional to the cluster
mass. The inferred 
correlation is consistent with a linear relation between the mass of the
central black hole(s) and the mass of the cluster, which is reminiscent of
the Magorrian relation between the black hole and bulge mass.

\end{abstract}


\section{Introduction}

Recent X-ray observations of both rich and poor clusters of galaxies have
shown that there are problems in understanding the temperature, gas density
or, equivalently, the entropy profiles at large radii ($\sim\,0.5r_{\rm
\scriptscriptstyle vir}$, where $r_{\rm \scriptscriptstyle vir}$ is the
virial radius) (Ponman \etal 2003). The observed entropy at $0.1r_{\rm
\scriptscriptstyle 200}$ and at $r_{\rm \scriptscriptstyle 500}$ (where
$r_{\rm \scriptscriptstyle 200}$  and $r_{\rm \scriptscriptstyle 500}$
correspond to radii within which the average overdensity is 200 and 500,
respectively) is higher than that estimated from the purely gravitational
interaction of gas with dark matter (Ponman \etal 2003 and references
therein; see also Roychowdhury \& Nath 2003). Earlier X-ray observations by
Lloyd-Davies \etal (2000) had shown that entropy at $0.1r_{\rm
\scriptscriptstyle 200}$  reached a ``floor'' for poor clusters and
groups. However, recent results  of Ponman \etal (2003) have shown that the
observed entropy is higher than the gravitational expectations for all
clusters with emission-weighted temperatures \lTr in the range 1
$\hbox{--}$ 10 keV. The entropy at the much larger radius of $r_{\rm
\scriptscriptstyle 500}$ is also found to be higher than expected from
purely gravitational processes. 

Many theoretical models have been proposed to explain this phenomenon,
including models that involve heat input from supernovae (Valageas \& Silk
1999; Wu, Nulsen \& Fabian 2000), gas cooling (Bryan 2000; Voit \& Bryan
2001; Muanwong \etal 2002; Wu \& Xue 2002; Dav\'e, Katz \& Weinberg 2002,
Tornatore \etal 2003), accretion shocks (Tozzi \& Norman 2001; Babul \etal
2002) and quasar outflows (Nath \& Roychowdhury 2002). 

Observations have also revealed the presence of X-ray deficient bubbles in
the inner regions of many cooling flow clusters, e.g., the Hydra A cluster
(McNamara \etal 2000), Abell 2052 (Blanton \etal 2001, 2003), Abell 2597
(McNamara \etal 2001), Abell 4059 (Heinz \etal 2002), Abell 2199 (Johnstone
\etal 2002), and others. These bubbles are characterized by low X-ray
emissivity, implying low density compared to the ambient medium.  In most
of these cases, the cavities are clearly coincident with the radio lobes of
the AGN in the cluster center. However, some clusters also exhibit cavities
with weak or undetectable radio emission (known as ``ghost bubbles'' or
``ghost cavities'') located far away from the cluster centers, like
Perseus (Fabian \etal 2000), MKW3s (Mazzotta \etal 2002), and Abell 2597
(McNamara \etal 2001). These bubbles are also believed to be filled with
relativistic plasma or buoyant gas deposited by jets from the central AGN,
and are thought to rise through the ICM subsonically due to buoyancy. As
they rise, radiative and adiabatic losses reduce the energy of the
relativistic plasma inside the bubbles, resulting in a very low radio
flux. The discovery of these bubbles and their detailed observational study
have stimulated theoretical studies of the impact of these bubbles on the
intracluster medium. 

The evolution of these bubbles has been studied extensively in connection
with the cooling flow catastrophe in the centers of clusters. It has been
found recently that simple cooling flow model of clusters, which predict
that the temperature of gas in the central regions of clusters should be
very low (much less than 1 keV), are mostly 
in conflict with the observed temperature profiles (Peterson \etal 2001;
Allen \etal 2001 and many others). Several authors have addressed whether
the absence of very 
cold gas (below $\sim$ 1 keV) can be explained by AGN energy input via
outflows and bubbles (see, e.g., Tabor \& Binney 1993, Binney \& Tabor 1995,
Ruszkowski \& Begelman 2002,
Churazov \etal 2002, Kaiser \& Binney 2003, and Br\"uggen 2003a for
semi-analytic models; and Quilis \etal 2001, Reynolds \etal 2002, Br\"uggen
2003b, and Ruszkowski, Br\"uggen \& Begelman 2004, Basson \& Alexander
2003, Omma et al. 2003a, Omma \& Binney 2003b for numerical simulations). 

In this paper, we explore the possibility of heating the intracluster gas
at large radii via the ``effervescent heating'' mechanism (Begelman 
2001, Ruszkowski \& Begelman 2002).  Since the excess entropy requirements
at different radii are different, the questions we set out to answer are
the following: 

\begin{itemize}
\item[1.] Is it possible to satisfy simultaneously the entropy observations
at two fiducial radii ($0.1r_{\rm \scriptscriptstyle 200}$ and $r_{\rm
\scriptscriptstyle 500}$) with a single central heating source for all
clusters? We explore the parameter space of time-averaged jet luminosity
\lLr and the time for which this activity continues, $t_{\rm
\scriptscriptstyle heat}$. 
\item[2.] How does the entropy profile of the cluster gas evolve with time
when it is being heated and cooled simultaneously? 
\end{itemize}

The paper is organized as follows. In \S~2 we describe our model, including
initial conditions, details of the effervescent heating mechanism, our
prescriptions for cooling and convection, and the simulation method. The
results and discussion are presented in \S\S~3 and 4, respectively.
We discuss caveats of our model in \S~5 and summarize our conclusions in 
\S~6.

\section {Model}

Our model of the ICM assumes that the thermal gas remains in hydrostatic
equilibrium as it is heated by buoyant radio bubbles, originating from the
central AGN, via the ``effervescent heating''  mechanism. Here, we begin
with a description of the background dark matter potential of the cluster. 

\subsection { Dark matter density profile}

We assume that the dark matter density profile of the cluster, $\rho_{\rm
\scriptscriptstyle dm}(r)$,  is described by a self-similar form as
suggested by many high resolution $N$-body simulations. The profile is
expressed in terms of a characteristic radius, $r_{\rm \scriptscriptstyle
s}$, by 
\be
\rho_{\rm \scriptscriptstyle dm}(r) = {\rho_s \over {x(1+x)^
{\rm \scriptscriptstyle 2}}}
\label{eq:dm}
\ee
(see, e.g., Komatsu \& Seljak 2002), where $\rho_{\rm \scriptscriptstyle
s}$ is a characteristic density and $x = r/r_{\rm \scriptscriptstyle
s}$. We define a dimensionless ``concentration parameter'' $c\equiv {r_{\rm
\scriptscriptstyle vir}/r_{\rm \scriptscriptstyle s}}$, where the virial
radius $r_{\rm \scriptscriptstyle vir}$ is calculated from the spherical
collapse model (Peebles 1980) assuming overdensity $\Delta_{\rm
\scriptscriptstyle c}(z=0) = 100$ (Komatsu \& Seljak 2002). 
The characteristic density $\rho{\rm \scriptscriptstyle s}$ is then given by 
\be
\rho_{\rm \scriptscriptstyle s} = c^3 {M_{\rm \scriptscriptstyle vir}
\over 4 \pi r_{\rm \scriptscriptstyle vir}^3} \Bigl [{\ln(1+c) - {c \over
(1+c)}\Big ]} ^{-1},
\label{eq:rhos}
\ee

\noindent
where $M_{\rm \scriptscriptstyle vir}$ is the cluster virial mass.  The
concentration parameter $c$ can be approximated by 

\be
c=9\Bigl ({M_{\rm \scriptscriptstyle vir} \over 1.5 \times 10^{13}h^{-1}
M_{\rm \scriptscriptstyle \odot}}\Bigr )^{-0.13},
\label{eq:cpfit}
\ee

\noindent
according to numerical simulations by Bullock \etal (2001) (we use $h=0.71$).

The above set of equations specifies the dark matter density profile of a
particular cluster as a function of its virial mass. Next, we turn our
attention to the density profile 
of the gas in hydrostatic equilibrium with this dark matter distribution. 

To compare our results with observations, we present our results in terms
of the radii $r_{\rm \scriptscriptstyle 200}$ and $r_{\rm
\scriptscriptstyle 500}$, where 
the overdensities are 200 and 500, respectively. 
Both radii are functions of the cluster mass. 
The ratio $r_{\rm \scriptscriptstyle 500}/r_{\rm \scriptscriptstyle
200}\approx 0.65-0.67$ for the range of masses we have considered.\\

\subsection {Initial configuration of intracluster gas}

We use the ``universal temperature profile'' (Loken \etal 2002) as the
initial temperature profile of gas in hydrostatic equilibrium 
to determine the initial density profile. This profile is different from
the commonly used self-similar profile, which assumes that the intracluster 
gas density distribution scales as the dark matter density profile with a
constant of proportionality $f_{\rm \scriptscriptstyle b}$ = 
${\Omega_{\rm \scriptscriptstyle b} \over \Omega_{\rm \scriptscriptstyle
m}}$ (i.e., $\rho_{\rm \scriptscriptstyle gas}=f_{\rm \scriptscriptstyle b} 
\rho_{\rm \scriptscriptstyle dm}$). We use the ``universal temperature
profile'' instead of the self-similar profile because the former is claimed
to be the result of gravitational processes alone. Moreover, it does not
have the unrealistic turnover of the temperature profile in the inner
regions of the cluster, which one finds in the 
self-similar profile (Roychowdhury \& Nath 2003).

The initial temperature profile (normalized by the emission-weighted
temperature $\langle T \rangle$), in the radial range 
$0.04 r_{\rm vir} \leq r \leq r_{\rm vir}$, is given by

\be
{T_{\rm \it \scriptscriptstyle o } (r)\over \langle T \rangle} =
{b \over (1+{r/a})^
{\rm \delta}}
\label{eq:temp}
\ee

\noindent
where $b = 1.33, \, a = r_{\rm \scriptscriptstyle vir}/1.5$, and $\delta =
1.6$. To determine the emission-weighted temperature from the cluster mass,
we use a relation that arises from adiabatic evolution of the gas in the
cluster. Afshordi \& Cen (2002) have shown that the observations by
Finoguenov \etal (2001) of the $M_{\rm \scriptscriptstyle 500}
\hbox{--}\langle T \rangle$ relation in clusters can be understood from
gravitational processes alone. We therefore use this empirical relation 
($M_{\rm \scriptscriptstyle 500} \hbox{--}\langle T \rangle$) derived by
Finoguenov \etal (2001): 

\be
M_{\rm \scriptscriptstyle 500}=2.64\times 10^{13}
\Bigl ( {k_{\rm \scriptscriptstyle b}
\langle T \rangle \over 1 \, {\rm keV}} \Bigr )^{1.78}
{\rm M}_{\rm \scriptscriptstyle \odot},
\label{eq:M-T}
\ee

\noindent
where $k_{\rm \scriptscriptstyle b}$ is the Boltzmann constant and $M_{\rm
\scriptscriptstyle 500}$ has been calculated self-consistently by 
taking the total mass within the radius where the over-density is $\delta \ge 500$.

The equation of hydrostatic equilibrium for gas in a cluster with
temperature $T_{\rm \it \scriptscriptstyle o }(r)$ 
and density $\rho_{\rm \scriptscriptstyle gas0}(r)$ is 

\be
{1 \over \rho_{\rm \scriptscriptstyle gas0}(r)}{d \over dr} (P_{\rm
\scriptscriptstyle gas0}(r)) = - {GM(\le r) \over
r^2} ,
\label{eq:hydro_a}
\ee

\noindent
where
$P_{\rm \scriptscriptstyle gas0} = ({\rho_{\rm \scriptscriptstyle gas0}
/{\mu m_{\rm \scriptscriptstyle p}}})k_{\rm \scriptscriptstyle b}T_{\rm
\scriptscriptstyle 0}$, $M( \le r)$ is the total mass inside radius $r$,
and $\mu$ and $m_{\rm \scriptscriptstyle p}$ denote the mean molecular
weight ($\mu \, = \, 0.59$) and the proton mass, respectively. The boundary
condition imposed on the solution is that the gas-fraction, $f_{\rm
\scriptscriptstyle gas}\equiv M_{\rm gas}/M_{\rm dm}=0.13333$, within
$r_{\rm \scriptscriptstyle 200}$ is universal, as recently found by Ettori
(2003) for a sample of low- and high-redshift clusters. 

\subsection{Heating, cooling and convection}

\subsubsection{Heating}

In the ``effervescent heating'' model, the central AGN is assumed to
inflate buoyant bubbles of relativistic plasma in the ICM (Begelman 2001,
Ruszkowski \& Begelman 2002). 
The timescale for the bubbles to cross the cluster (of order the free-fall
time) is shorter than the cooling timescale. Since the number flux of
bubbles is large, the flux of bubble energy through the ICM approaches a
steady state. This implies that details of the energy injection process
such as the number flux of bubbles, the bubble 
radius or size, filling factor and rate of rise do not affect the average
heating rate. 

It is assumed that the relativistic gas does not mix with the ICM very
efficiently. Under such conditions the bubbles can expand and do $pdV$ work
on the ambient medium as they rise in the cluster pressure gradient. We
assume that this work converts internal energy of the bubbles to thermal
energy of the intracluster gas within a pressure 
scale height of where it is generated. Then, in a steady state (and
assuming spherical symmetry), the energy flux carried by the bubbles varies
as

\be
F_{\rm \scriptscriptstyle b} \propto \frac{P_{\rm \scriptscriptstyle b}(r)^{(\gamma_{\rm
\scriptscriptstyle b}-1)/\gamma_{\rm \scriptscriptstyle b}}}{r^{2}}
\ee

\noindent
where $P_{\rm \scriptscriptstyle b}(r)$ is the partial pressure of buoyant
gas inside the bubbles at radius $r$ and $\gamma_{\rm \scriptscriptstyle
b}$ is the adiabatic index of buoyant gas, which we have taken to be 4/3. 
This formula includes the effects of adiabatic bubble inflation.
Assuming that the partial pressure inside these bubbles scales as the
thermal pressure of the ICM, the volume heating 
function ${\cal H}$ can be expressed as

\bea
{\cal H} &\sim& -r^{2}h(r) {\mathbf {\nabla}}\cdot ({\mathbf {\hat
r}}F_{\rm \scriptscriptstyle b})\nonumber \\
&=& -h(r)\Bigl ({P_{\rm \scriptscriptstyle gas} \over P_{\rm
\scriptscriptstyle 0}} \Bigr )^{
(\gamma_{\rm \scriptscriptstyle b}-1)/\gamma_{\rm \scriptscriptstyle b}}
{1 \over r}{{d \ln P_{\rm \scriptscriptstyle gas}} \over {d \ln r}},
\label{eq:heatingrate}
\eea

\noindent
where $P_{\rm \scriptscriptstyle 0}$ is some reference pressure and $h(r)$
is the normalization function 

\be
h(r) = {{\langle L \rangle} \over {4 \pi r^2}}(1-\exp(-r/r_
{\rm \scriptscriptstyle 0})) \, q^{-1} .
\ee

\noindent
In equation (9), \lLr is the time-averaged energy injection rate and
$r_{\rm \scriptscriptstyle 0}$ is the inner heating cut-off radius which is
determined by the size of the central heating source. In our calculations
$r_{\rm \scriptscriptstyle 0}$ is taken to be $0.01r_{\rm
\scriptscriptstyle vir}$, which is $\sim$ 15 $\hbox{--}$ 20 
kpc depending on the cluster mass $M_{\rm \scriptscriptstyle cl}(\equiv
M_{\rm vir})$.  The normalization factor $q$ is defined by 

\be
q=\int_{r_{\rm \scriptscriptstyle min}}^{r_{\rm \scriptscriptstyle max}}
\Bigl ({P \over P_{\rm \scriptscriptstyle 0}} \Bigr )^{(\gamma_{\rm
\scriptscriptstyle b}-1)
/\gamma_{\rm \scriptscriptstyle b}}{1 \over r}{{d \ln P} \over {d \ln r}}
(1-e^{-r/r_{\rm \scriptscriptstyle 0}})\, dr ,
\ee
where $r_{\rm{max}}=r_{200}$.

\noindent

\subsubsection{Cooling and convection}

To calculate the volume cooling rate, we use a fit to the normalized
cooling function $\Lambda_{\rm \scriptscriptstyle N}(T)$ for a metallicity
of $Z/Z_{\rm \scriptscriptstyle \odot}$ = 0.3, as calculated by Sutherland
\& Dopita (1993). This cooling function incorporates the effects of
free-free emission and line cooling. The fit is borrowed from Nath
(2003). Thus, the volume cooling rate is $\Gamma = n_{\rm
\scriptscriptstyle e}^2 \Lambda_{\rm \scriptscriptstyle N}(T)$, where
$n_{\rm \scriptscriptstyle e}$ = $0.875 (\rho/m_{\rm \scriptscriptstyle
p})$ is the electron density. 

The convective flux $F_{\rm \scriptscriptstyle conv}$ is given by the mixing length theory,

\be
F_{\rm \scriptscriptstyle conv} = \cases{{2^{-5/2}
c_{\rm \scriptscriptstyle p}^{-1/2}}Tg^{1/2}\rho_{\rm
\scriptscriptstyle gas}l_{\rm \scriptscriptstyle m}^{2}(-\nabla\hat{s})
^{3/2}&if $\nabla\hat{s}<0$,\cr
0&otherwise,\cr}
\label{eq:conv}
\ee

\noindent
where $g$ is the gravitational acceleration, $l_{\rm \scriptscriptstyle m}$
is the mixing length, $\hat{s} = (\gamma -1)^{-1}k_{\rm \scriptscriptstyle
b}/(\mu m_{\rm \scriptscriptstyle p})\ln(P_{\rm \scriptscriptstyle
gas}/\rho_{\rm \scriptscriptstyle gas}^{\gamma})$ is the gas entropy per
unit mass, and $c_{\rm 
\scriptscriptstyle p}=\gamma k_{\rm \scriptscriptstyle b}/[(\gamma -1)\mu
m_{\rm \scriptscriptstyle p}]$ is the specific heat per unit mass at
constant pressure. We use $l_{\rm \scriptscriptstyle m} = \min [0.3P_{\rm
\scriptscriptstyle gas}/(\rho_{\rm \scriptscriptstyle gas}g),r]$, where $r$
is the distance from the cluster center. 

\subsubsection{Evolution of the intracluster gas}

As noted earlier, the gas is assumed to be in quasi-hydrostatic equilibrium
at all times since the cooling is not precipitous at these radii and the
heating is mild. The gas entropy per particle is

\be
S = const + \frac{1}{\gamma -1}k_{\rm
\scriptscriptstyle b}\ln(\sigma).
\ee

\noindent
where $\sigma \equiv P_{\rm \scriptscriptstyle gas}/\rho_{\rm
\scriptscriptstyle gas}^{\rm \scriptscriptstyle \gamma}$ is the ``entropy
index'' and $\gamma$ is the 
adiabatic index. The particle number density of the gas, $n$, is given by
$n = \rho_{\rm \scriptscriptstyle gas}/\mu m_{\rm 
\scriptscriptstyle p}$.

During each timestep $\Delta t$, the entropy of a given mass shell changes by an amount

\be
\Delta S = \frac{1}{\gamma -1} k_{\rm \scriptscriptstyle b} {\Delta \sigma \over
\sigma} = {1 \over {n T}}({\cal H} - \Gamma -
\nabla\cdot F_{\rm \scriptscriptstyle conv})\Delta t .
\ee

\noindent
Incorporating the expressions for heating, cooling, and convection, the
entropy increment for each mass shell for a timestep $\Delta t$ becomes 

\bea
\Delta \sigma (M)= {2 \over 3} {\sigma \over P_{\rm \scriptscriptstyle gas}}
\Delta t \left[ -h(r) \Bigl
({P_{\rm \scriptscriptstyle gas} \over P_{\rm \scriptscriptstyle 0}} \Bigr
)^{(\gamma_{\rm \scriptscriptstyle b}-1)/\gamma_{\rm \scriptscriptstyle b}}
\nonumber \right.\\
\times {1 \over r}{{d \ln P_{\rm \scriptscriptstyle gas}} \over {d \ln r}} \, -
n_{\rm \scriptscriptstyle e} ^2\Lambda_{\rm \scriptscriptstyle
N}(T) \nonumber \\
\left. -{1 \over {r^2}}{d \over dr}(r^2 F_{\rm \scriptscriptstyle conv})
\right]
\label{eq:del_sig}
\eea

\noindent
where $n_{\rm \scriptscriptstyle e}$ and $P_{\rm \scriptscriptstyle gas}$
are the current electron number density and pressure of the ICM,
respectively. 

Thus, the entropy index of each mass shell of gas due to heating and
cooling after a time $\Delta t$ becomes 

\be
\sigma_{\rm \scriptscriptstyle new}(M) = \sigma_{\rm \scriptscriptstyle 0}
(M) + \Delta \sigma (M)
\label{eq:sig_update}
\ee

\noindent
where $ \sigma_{\rm \scriptscriptstyle 0}(M) = P_{\rm \scriptscriptstyle
gas0}/\rho_{\rm \scriptscriptstyle gas0}^{\rm \scriptscriptstyle \gamma}$
is the default entropy index. The system relaxes to a new state of
hydrostatic equilibrium with a new density and temperature profile. After
updating the function $\sigma (M)$ for each mass shell, we solve the
equations 
\bea
{dP_{\rm \scriptscriptstyle gas} \over dM}&=& {G M(\le r) \over
{4\pi r^{\rm \scriptscriptstyle 4}}}\label{eq:hydro_b}
\\
\, {dr \over dM} &=& {1 \over 4\pi r^{\rm \scriptscriptstyle 2}}
{\Bigl({P_{\rm \scriptscriptstyle gas}(M) \over \sigma(M) }\Bigr )}^{(\rm
\scriptscriptstyle 1/\gamma)}
\label{eq:evol_gas}
\eea
to determine the new density and temperature profiles at time $t+\Delta t$.
The boundary conditions imposed on these equations are that (1) the
pressure at the boundary of the cluster, $r_{\rm \scriptscriptstyle out}$,
is constant and is equal to the initial pressure at $r_{\rm
\scriptscriptstyle 200}$, i.e., $P\,(r_{\rm \scriptscriptstyle out}$) =
constant = $P_{\rm \scriptscriptstyle gas0}$($r_{\rm \scriptscriptstyle
200}$), and (2) the gas mass within $r_{\rm \scriptscriptstyle out}$ at all
times is the mass contained within $r_{\rm \scriptscriptstyle 200}$ for the
default profile at the initial time, i.e., $M_{\rm \scriptscriptstyle g}
(r_{\rm \scriptscriptstyle out}) = M_{\rm \scriptscriptstyle g0}(r_{\rm
\scriptscriptstyle 200}) = 0.1333M_{\rm \scriptscriptstyle dm} (r_{\rm
\scriptscriptstyle 200})$. It is important to note here that 
$r_{\rm \scriptscriptstyle out}$ increases as the cluster gas gets heated and spreads out.

The observed gas entropy ${\cal S}$($r$) at $0.1r_{\rm \scriptscriptstyle
200}$ and at $r_{\rm \scriptscriptstyle 500}$ is then calculated using 

\be
{\cal S}(r)\equiv T(r)/n_{\rm \scriptscriptstyle e}^{2/3}(r) .
\ee

\noindent
The updated values of $\sigma (M)$ and pressure of the ICM $P_{\rm
\scriptscriptstyle gas}(r)$ are used to calculate the heating and cooling
rates and the convective flux for the next time step. This is continued for
a duration of $t_{\rm \scriptscriptstyle heat}$.  After that the heating
source is switched off, putting ${\cal H}$ = 0. 
The cooling rate and convective flux continue to be calculated to update
the function $\sigma (M)$ at subsequent timesteps, and the hydrostatic
structure is correspondingly evolved for a duration of $t_{\rm
\scriptscriptstyle H} - t_{\rm \scriptscriptstyle heat}$, where $t_{\rm
\scriptscriptstyle H}$ is the Hubble time. Note that $r_{\rm
\scriptscriptstyle out}$ decreases during this time since the intracluster
gas loses entropy and shrinks. 
The only free parameters in our calculation are the energy injection rate 
\lLr and the time $t_{\rm \scriptscriptstyle heat}$ over which the
``effervescent heating'' of the ICM takes place.
After evolving the gas for the total available time, $t_{\rm
\scriptscriptstyle H} \sim 1.35\times 10^{10}$ years, we check whether the
entropy at $0.1r_{\rm \scriptscriptstyle 200}$ and $r_{\rm
\scriptscriptstyle 500}$ matches the observed values, and adjust parameters
accordingly. In this way, we explore the parameter space of \lLr and
$t_{\rm \scriptscriptstyle heat}$ for different cluster masses so that the
entropy (after 1.35$\times$10$^{10}$ years) at $0.1r_{\rm
\scriptscriptstyle 200}$ and $r_{\rm \scriptscriptstyle 500}$ matches the
observed values. 

For numerical stability of the code, the convection term is integrated
using timesteps that satisfy the appropriate Courant condition. The Courant
condition for convection is 

\be
\Delta t_{\rm \scriptscriptstyle conv} \le {1 \over 2} 2^{5/2} \sqrt{\gamma}
{(\Delta r)^{5/2} \over {g^{1/2}l_{\rm \scriptscriptstyle m}^2}} .
\ee

\noindent
The timesteps, $\Delta t$,  used in equation (14) to update the entropy of
the gas and calculate its pressure, temperature and density 
profiles always obey the above Courant condition (Ruszkowski \& Begelman
2002; Stone, Pringle \& Begelman 1999). 

The calculations presented in this paper were also performed using the
fully time-dependent {\it ZEUS} code (Clarke, Norman \& Fiedler 1994) and
the results obtained were consistent with the ones presented here. However,
we decided to use our quasi-hydrostatic Lagrangian code as it allowed us to
search the parameter space more efficiently. 

\section{Results}

In this section, we discuss our results for cluster evolution due to
heating, cooling and convection. The gas is heated for a time $t_{\rm
\scriptscriptstyle heat}$ and cooled simultaneously. After this time, the
heating source is switched off. The gas is then allowed to cool radiatively
until a total simulation time of $t_{\rm \scriptscriptstyle
H}$=1.35$\times$10$^{10}$ years has elapsed. The final entropy values at
$0.1r_{\rm \scriptscriptstyle 200}$ and $r_{\rm \scriptscriptstyle 500}$
are compared with the observed ones. 

In Figure 1, the observed entropy values and their errors at
$0.1r_{\rm \scriptscriptstyle 200}$ and at $r_{\rm
\scriptscriptstyle 500}$ are plotted as a function of the
emission-weighted temperature, \lTr, of the cluster (Ponman \etal
2003). We have done a best-fit analysis on these data points to
estimate the entropy requirements for the sample of clusters of
masses ranging from $10^{14}$ to $2 \times 10^{15}$ $M_{\rm
\scriptscriptstyle \odot}$. The results of our analysis are the
shaded regions in the two panels of Figure 1. The solid line
through the center of the shaded region is the best-fit curve with
the lines bounding the shaded region being the $1-\sigma$ errors
on the best-fit entropy values.

Figure 2 shows the time evolution of scaled entropy profiles of a
cluster of mass $M_{\rm \scriptscriptstyle cl} = 6\times10^{14}
M_{\rm \scriptscriptstyle \odot}$ with and without convection for
\lLr = 3$\times 10^{45}$ erg s$^{-1}$. We use the same method of
emissivity weighting as in Roychowdhury \& Nath (2003) to calculate the
average quantities. The entropy profiles are plotted in time-steps
of $5\times10^{8}$ years. They are seen to rise with time as the
ICM is heated. Then, after the heating is switched off (after
$t_{\rm \scriptscriptstyle heat} = 5\times 10^9$ years), the gas
loses entropy due to cooling and the profiles are seen to fall
progressively until the Hubble time is reached.  The inclusion of
convection flattens the negative gradient in the scaled entropy
profiles in the central regions of the cluster (within $0.2r_{\rm
\scriptscriptstyle 200}$, as seen in the difference of the entropy
profiles between the left and right panels in Figure 2. The plot
that includes convection (left panel in Figure 2) shows that the
gas develops a flat entropy core in its central regions after the
heating source has been switched off and the gas has cooled.

Figure 3 shows the evolution of density and temperature of the ICM
for a cluster of mass $6\times10^{14}$ $M_{\rm \scriptscriptstyle
\odot}$ and for a luminosity of \lLr $ = 3\times 10^{45}$ erg
s$^{-1}$.  The gas density decreases with time during the heating
epoch, and increases due to radiative cooling and convection after
the heating source is switched off. It is interesting to note that
the changes in density are minimal beyond $0.2r_{\rm
\scriptscriptstyle 200}$, and that convection plays an important
role in regulating the density profiles after the heating source
is switched off for radii $r \le 0.2r_{\rm \scriptscriptstyle
200}$. We note that for clusters with lower emission-weighted
temperatures, the effects of heating and convection are seen at
larger radii.

We now discuss the permitted range in the total energy injected
into the cluster, $E_{\rm \scriptscriptstyle agn} =$ \lLr
$\times\, t_{\rm \scriptscriptstyle heat}$, required to match the
observed entropy as a function of the cluster mass. Figure 4 shows
the spread in $E_{\rm \scriptscriptstyle agn}$ as a function of
the cluster mass for two different values of $t_{\rm
\scriptscriptstyle heat}$. The region bounded by thin solid lines
is the permitted range in energy for $t_{\rm \scriptscriptstyle
heat}$ = $5\times 10^{8}$ years. This region includes an area shaded with
solid vertical lines, which corresponds to the values of $E_{\rm
\scriptscriptstyle agn}$ that satisfy the entropy requirement at
$0.1r_{\rm \scriptscriptstyle 200}$, and another area shaded with
solid oblique lines which satisfies the entropy requirement at
$r_{\rm \scriptscriptstyle 500}$. Similarly, the region bounded by
thin dotted lines in Figure 4a shows the permitted spread in energy for $t_{\rm
\scriptscriptstyle heat}= 5\times10^{9}$ years. This region
includes an area shaded with dots which corresponds to the values
of $E_{\rm \scriptscriptstyle agn}$ that satisfy the entropy
requirement at $0.1r_{\rm \scriptscriptstyle 200}$ and another
area shaded with long-dashed horizontal lines, which satisfies the
entropy requirement at $r_{\rm \scriptscriptstyle 500}$. 
For cluster masses above $9\times 10^{14}M_{\odot}$ there is no lower limit
on the injected energy from the entropy measurements at $r_{500}$. The region
corresponding to this situation is marked by
by horizontal dashed lines and oblique lines for cluster masses above
$9\times 10^{14}M_{\odot}$.

Figure 5 shows the permitted total injected  energy range as a
function of the mass of cluster for heating times between 
$t_{\rm heat}=5\times 10^{8}$ years and
$t_{\rm heat}=5\times 10^{9}$ years. Here the entropy is required to
match observations at {\it both} $0.1r_{\rm \scriptscriptstyle 200}$ and
$r_{\rm \scriptscriptstyle 500}$. The thick solid line represents a linear
relation between the total energy  injected to the cluster by AGN and
the mass of the cluster (see next section for more details).

\section{Discussion}

As is clear from Figures 4 and 5, it is possible to heat the ICM
with a {\it single} central AGN to match the entropy requirements
at {\it both} $0.1r_{\rm \scriptscriptstyle 200}$ and $r_{\rm
\scriptscriptstyle 500}$. However, in order to match the entropy
at both radii, the total injected energy $E_{\rm
\scriptscriptstyle agn}$, 
for a given value of $t_{\rm \scriptscriptstyle heat}$ $\ll$
$t_{\rm \scriptscriptstyle H}$, must be tightly constrained.  In
fact, our calculations have shown that for any value of $t_{\rm
\scriptscriptstyle heat} < t_{\rm \scriptscriptstyle H}$, i.e.,
for any heating time (or AGN lifetime), it is always possible to satisfy the entropy
observations at {\it both} radii with a {\it single} value of the
luminosity \lLr. This is different from the cooling flow problem,
where \lLr must be finely tuned to match the cooling rate
(Ruszkowski \& Begelman 2002), because cooling effects on large
scales are rather mild and, thus, the results depend mostly on the
total injected energy $E_{\rm \scriptscriptstyle agn}=\langle
L\rangle\,\times\,t_{\rm heat}$.  Thus, if we can fit the observed
entropy values for just one pair \lLr and $t_{\rm heat}$, we can
do so for a wide range of such pairs. Another manifestation of
this fact is that the plots in Figure 4 are similar in shape but
are just offset by a factor of a few. 
As the cooling effects are relatively mild, 
the modest differences arise because shorter heating times lead to higher
temperatures and convective transport 
cannot ``catch up'' with the energy supply. As the entropy excess to
be explained is known from observations and is fixed, more heat has to be injected 
for shorter heating times.
Nevertheless, the results are mostly sensitive to the {\it
total} energy input from the black hole, rather than to \lLr and
$t_{\rm heat}$ separately.  As a consequence, satisfactory fits
can be obtained as long as the total injected energy falls within
a relatively narrow range of values, which depends on the cluster
mass (Figure 5).

As an illustrative example of such a relation, the thick solid
line plotted in Figure 5 shows a linear relation between $E_{\rm
\scriptscriptstyle agn}$ and cluster mass, $M_{\rm
\scriptscriptstyle cl}$, that is consistent with the requirements
from entropy observations for a heating time of
$5\times 10^8\le t_{\rm \scriptscriptstyle heat}\le 5\times 10^9$ years.
We point out that the effective heating time may be longer than the
integrated AGN lifetimes. Heating of the cluster may occur more 
gradually as heat gets distributed on a timescale very roughly
proportional to the sound crossing time
from the cluster center to a given radius. Thus, 
$t_{\rm \scriptscriptstyle heat}$ should be interpreted as an upper limit on
the AGN lifetime.  

Let us assume
that the bulge mass of the central AGN is related by a factor $f$
to the cluster mass as $M_{\rm \scriptscriptstyle h} \sim f \,
M_{\rm \scriptscriptstyle cl}$ and that the mass of the black hole
is related to the bulge mass of the AGN host by $M_{\rm
\scriptscriptstyle bh} \sim 1.5\times 10^{-3} M_{\rm
\scriptscriptstyle h}$ 
(H\"aring \& Rix 2004; Ferrarese \& Merritt 2000, Gebhardt \etal
2000).  This means that the mass of the black hole is related to
the cluster mass by $M_{\rm bh}\sim 1.5\times 10^{-5}f_{-2}M_{\rm
cl}$, where $f_{-2}=f/0.01$. It is generally
thought that AGN are fuelled by accretion onto the
supermassive black hole. If we require that a fraction
$\eta$ of the mass of the black hole is converted into energy, 
then the relation between the black hole mass and the total
injected energy becomes $M_{\rm bh}\sim 2.8\times
10^{8}\eta_{0.2}^{-1}E_{62}{\rm M}_{\odot}$, where
$\eta_{0.2}=\eta/0.2$.
The relation between the total injected energy and the cluster
mass is $E_{62}\sim 5.4\eta_{0.2}f_{-2}M_{14}$. 
The linear relation in Figure 5 corresponds to 
$f_{-2}\eta_{0.2}=0.25$.

Note that for longer heating times the luminosity constraints become less
stringent and lower black hole masses can explain the observed
trends. Finally, as we consider heating on large scales, all
galaxies in the cluster that go through an active phase will
contribute to the overall energy budget of the cluster gas. For
example, substructure in the cluster could contain galaxies with
sizable bulge components, each of which may contain a supermassive
black hole. Therefore, the constraint on the black hole mass
obtained above should be interpreted as a sum of all black hole
masses that contribute to heating rather than the mass of an
individual black hole. This could also lower the required efficiency of individual
black holes that contribute to cluster heating.

Finally, we note that cooling and convection play important roles in
controlling the heating mechanism so that the entropy profiles broadly
match the observed entropy profiles in clusters (Ponman \etal
2003). Notably, in the later stages of evolution of the gas, after the
heating source is switched off, convection flattens the entropy profiles in
the central regions of the cluster. Moreover, an entropy core seems to
develop in the final stages of the evolution of the ICM. This is consistent
with the observed entropy profiles, which show cores at $r \,\le\,
0.1r_{\rm \scriptscriptstyle 200}$ (Ponman \etal 2003).  Our entropy
profiles do not show steep positive gradients as observed in the entropy
profiles of two groups of galaxies (Mushotzky \etal 2003). However, these
groups have masses smaller than the ones considered here. Moreover,
since the temperature profiles show a temperature gradient even though the
entropy is flattened, thermal conduction might be able to conduct heat out from the
central regions of the cluster and reduce temperatures there. This will happen at
roughly constant pressure and, therefore, central densities will slightly increase.
Thus conduction (if not fully suppressed) may help to remove the core in
the center. Analyzing the effects of conduction 
is not within the scope of this paper.

\section{Caveats}
In this section we make explicit the caveats of our model.
We note that our model assumes that all of the injected energy
goes into effervescent heating.  The amount of energy that is
available to effervescent heating, as opposed to the energy that is
``lost'' to bubble creation, can be estimated from the following
simple analytic argument.          The available effervescent energy
equals the energy lost to bubble expansion as it rises buoyantly,
$\int_{P_{o}}^{0}PdV$, where $P_{o}$ is the pressure at the transition
region from the bubble formation region to the buoyant (effervescent)
phase.  To get the maximum energy available we set the upper limit of
the integral to zero pressure. Assuming adiabatic evolution of the gas
inside the bubbles (low density, low radiative losses) and mass
conservation in the bubble one has
$dV=(1/\gamma)(P_{o}/P)^{1/\gamma}V_{o} dP/P$, where $V_{o}$ is the
volume of the bubbles at the transition region mentioned above. On
integrating, effervescent energy is $3P_{o}V_{o}$ for $\gamma=4/3$
(appropriate to the bubble interior). Thus, since the energy fraction 
lost to bubble
creation at constant pressure is just $p_{o}V_{o}$,  approximately
only $25\%$ of the injected energy goes to bubble creation, while
$75\%$ is available for effervescent heating. We note that if bubble
creation occurs at constant pressure, then the injected energy is
converted to $pdV$ work on the ICM. The exact details of the
dissipation of this energy are as yet unknown, but presumably this
process occurs with efficiency lower than $100\%$. Thus, in the regime
where bubble inflation occurs at constant pressure, the above estimate
of energy lost to the bubble creation is probably a lower limit.
However, if the bubble inflation in the cluster center occurs
supersonically, the energy required to inflate the bubble  goes into
shock heating. Any shocked, high entropy gas is buoyant itself, and
acts in a fashion similar to the case described by the effervescent
heating function the difference being that  a small fraction
of the total mass from the innermost radial shells is removed.\\
\indent
We note that our treatment is simplified as we neglect cosmological
evolution. In the hierarchical
picture, temperature increases with merging as small objects merge
to form large objects with higher masses.
Also, merging can only increase the entropy. Therefore, heating
of the smaller units before they merge, and therefore when they have lower
virial temperatures, should reduce the total energy requirement 
according to the second law of thermodynamics ($dE=TdS$). 
As clusters observed today ($z\sim 0$) were most likely formed at 
$z\la 0.5$ (Kitayama \& Suto 1996, Balogh, Babul \& Patton 1999)
and AGN could switch on before the formation epoch of these clusters,  
lower energy injection could be
required to explain observations at $z\sim 0$.\\
\indent
We also note that the hydrodynamics equations have been integrated
up to $r_{200}$. It is likely that the buoyant energy 
transport would be inhibited by the cluster accretion shock at such
large radii. However, excess entropy used to constrain our model 
is measured at smaller radii and we do not expect this to affect our
results significantly.\\
\indent
Our simulations assume that heating is instantaneous.  This is an
approximation as there should be a delay between the onset of AGN
activity and heating at a given radius.  We have chosen the shortest
heating duration to be  at least the sound crossing time (or the
dynamical time of rising bubbles) at the outer radius where we compare
the computed and observed entropies ($r_{500}$). Thus, for the
shortest heating time, our assumption at $0.1r_{200}$ should be quite
reasonable but at $r_{500}$ it would be less accurate but still
physically conceivable. For longer heating times our results at both
radii should not be affected by this approximation.\\
\indent
Related is the issue of the constraints on the source luminosity.
For shorter heating times, higher luminosities are required to assure
the same energy injection. Thus, the required luminosities are more
feasible for longer heating times. We note also that if the heating is
supplied by more than one AGN, then the luminosity requirements on an
individual source would be lower. We also reiterate the point made
earlier that  if the effects of cosmological evolution had been fully
taken into account and AGN had supplied heating at earlier epochs,
then the energy requirements presented here would have been reduced.
Thus, the source luminosities would also be lower for a given
heating time. This is due to the fact that the required excess entropy
at higher redshift would be reduced.

\section{Conclusions}

We have studied whether ``effervescent heating'' of cluster gas by a
central AGN can resolve the entropy problem in clusters of galaxies.  In
this model, the AGN (or a group of AGNs in the central region) injects
bubbles of buoyant gas, which heat the ICM. The mean volume heating rate
due the bubbles is a function of the ambient pressure and a time-averaged
energy injection rate to the ICM, but not of the detailed properties of the bubbles or
their evolution.  We have also included the effects of radiative cooling
and convection.  We assumed that heating continues for $t_{\rm
\scriptscriptstyle heat}$, the duration of heating, and have studied the
resulting evolution of the gas assuming it to be in quasi-hydrostatic
equilibrium for the Hubble time. The only free parameters of this model are
\lLr, the energy injection rate to the ICM, and $t_{\rm \scriptscriptstyle
heat}$. The main results of our study are summarized as follows: 

\begin{itemize}
\item[1.] We find that there are allowed values of \lLr for which it is
possible to match the entropy observations of clusters even at large radii
(both $0.1r_{\rm \scriptscriptstyle 200}$ and $r_{\rm \scriptscriptstyle
500}$) with a {\it single} central AGN for a large range of $t_{\rm
\scriptscriptstyle heat}$ ($5\,\times\,10^8\,\le\, 
t_{\rm \scriptscriptstyle heat}\,\le\,t_{\rm \scriptscriptstyle H}$ years)
and cluster masses ($5\times 10^{13}-2\times 10^{15}{\rm M_{\odot}}$).
 
\item[2.] Convection plays an important role in removing negative entropy
gradients produced by heating in the central regions of the cluster.
 
\item[3.]
The results are mostly sensitive to the total energy $E_{\rm agn}$ injected
into the cluster by AGN (as cooling effects are relatively mild).  The
model predicts that the total injected energy $E_{\rm
agn}=$\lLr$\times\Delta t$, required to satisfy observational entropy
constraints, should be correlated with the cluster mass. This requirement
is consistent with a linear relation between the mass of the central black
hole(s) and the mass of the cluster, which is reminiscent of the Magorrian
relation between the mass of the black hole and the bulge mass of the host
galaxy. 

\end{itemize}

\acknowledgments

We are grateful to the anonymous referee for his/her very
constructive comments which helped to improve the paper. We also thank 
James Binney for sharing with us his views on the details of AGN heating.
SR and BBN would like to thank their colleagues in Raman Research Institute
for stimulating discussions.  MR thanks Phil Armitage for discussions and
frequent tea breaks. 
Support for this work was provided by National Science Foundation grant
AST-0307502 and the National Aeronautics and Space Administration through
{\it Chandra} Fellowship Award Number PF3-40029 issued by the {\it Chandra}
X-ray Observatory Center, which is operated by the Smithsonian
Astrophysical Observatory for and on behalf of the NASA under contract
NAS8-39073.

\clearpage

\begin{figure}
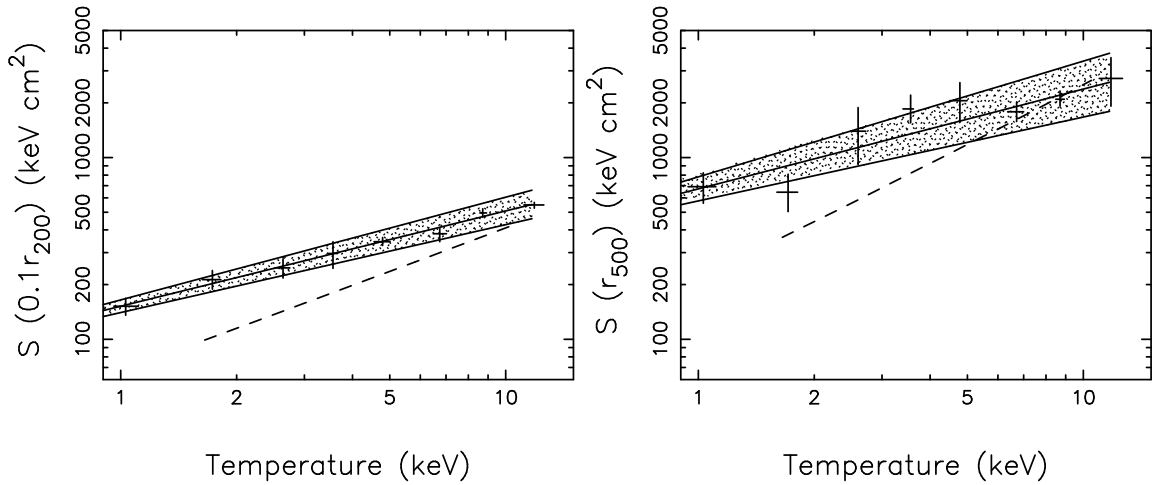

\includegraphics[width=2.5in,angle=270]{f1a.eps}
\includegraphics[width=2.5in,angle=270]{f1b.eps}
\caption{Gas entropy (defined here as $T/n_{e}^{2/3}$) as a function of
emission-weighted temperature \lTr, 
at radii $0.1r_{\rm \scriptscriptstyle 200}$ (left panel) and $r_{\rm
\scriptscriptstyle 500}$ (right panel). The data points are from Ponman
\etal (2003). The solid line in the center of the shaded region is the
best-fit to the data points. The two solid lines bounding the shaded region
are the $1-\sigma$ errors on the best-fit values of entropy. The dashed
line is the predicted entropy due to gravitational interactions alone (from
Roychowdhury \& Nath 2003).} 
\label{fig:entropy}
\end{figure}

\clearpage

\begin{figure}
\centering
\includegraphics[width=3.0in]{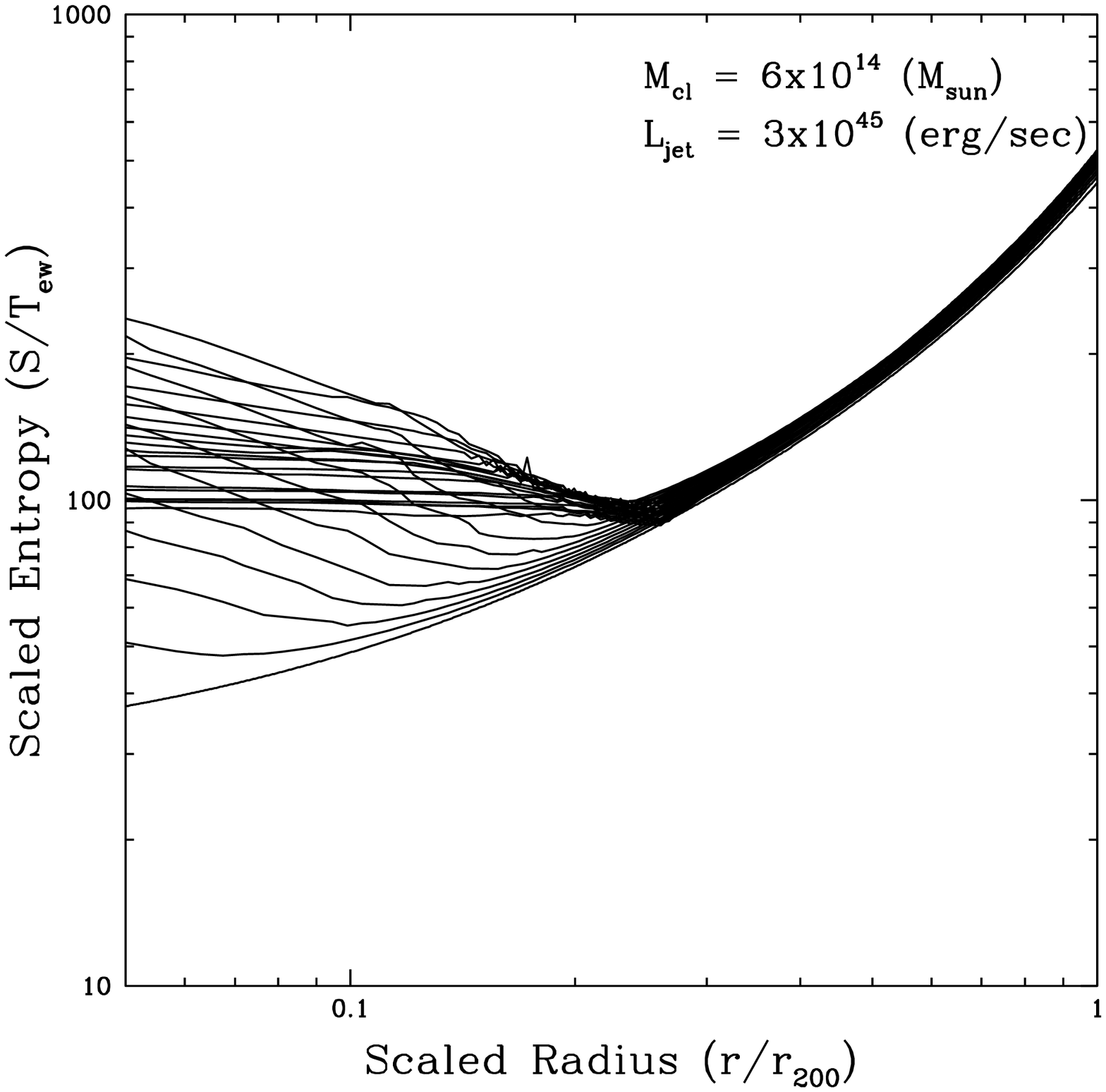}
\includegraphics[width=3.0in]{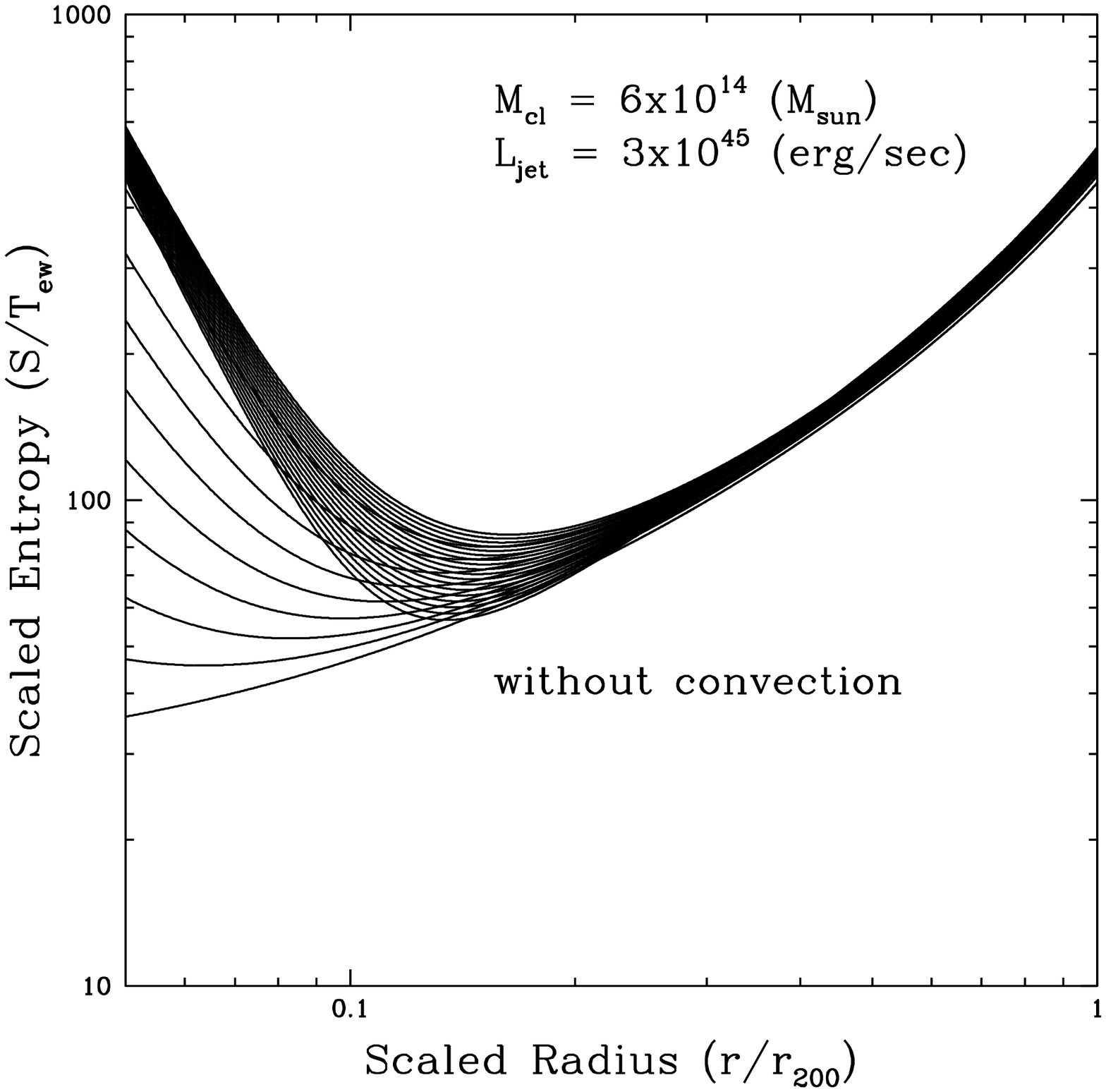}
\caption{Scaled entropy profiles as a function of scaled radius for a
cluster of mass $6 \times 10^{14}$ $M_{\rm \scriptscriptstyle \odot}$
heated by an AGN with \lLr$=3\times 10^{45}$ erg s$^{-1}$, with convection
(left panel) and without convection (right panel). The scaled entropy
profiles are plotted at intervals of 5$\times$10$^8$ years. They are seen
to rise as the gas is heated and then fall as the gas cools. In both cases
$t_{\rm \scriptscriptstyle heat} = 5\times 10^9$ years.
Initial states correspond to the lowest curves.}
\label{fig:sc_ent}
\end{figure}

\clearpage

\begin{figure}
\centering
\includegraphics[width=3.0in]{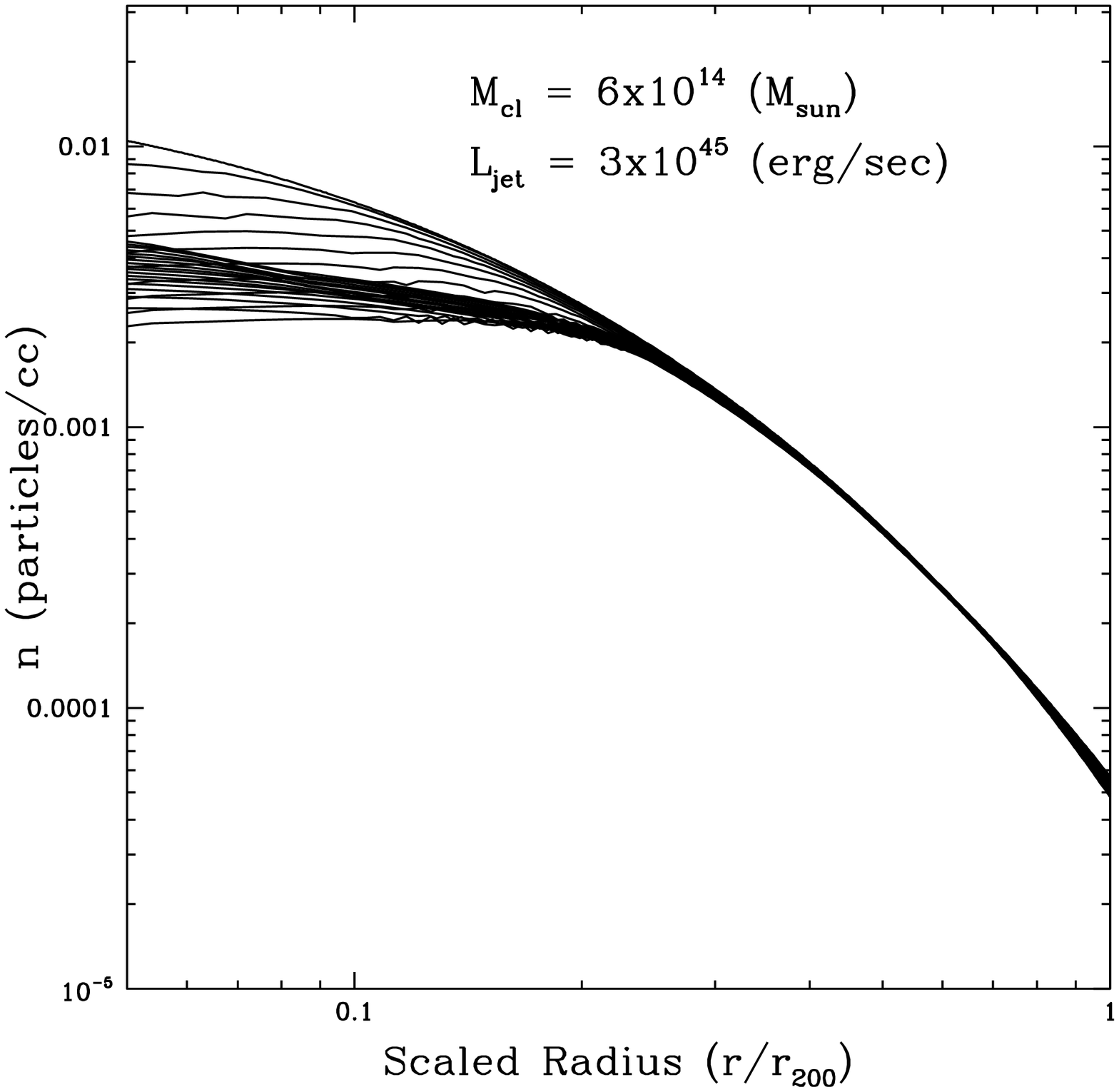}
\includegraphics[width=3.0in]{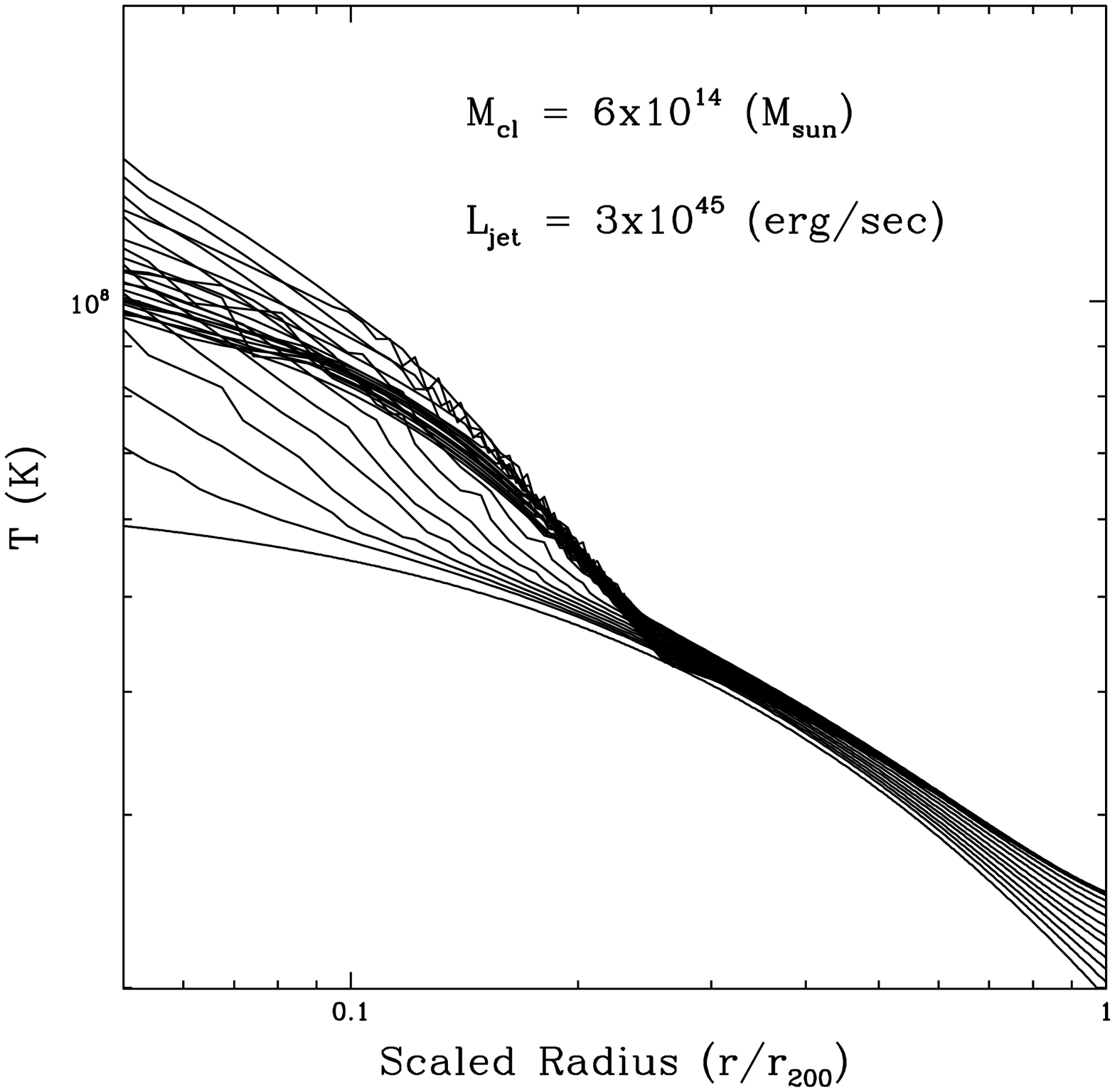}
\caption{Gas density (left panel) and temperature (right panel) profiles
as a function of scaled radius ($r/r_{\rm \scriptscriptstyle 200}$), for
the cluster model shown in the left panel of Figure 2.  It is seen here
that radiative cooling lowers the temperature, thus increasing the density
after the heating source is switched off. Initial density and temperature
profiles correspond to the highest and lowest curves, respectively.}
\label{fig:gas614}
\end{figure}

\clearpage

\begin{figure}
\centering
\includegraphics[width=3.0in]{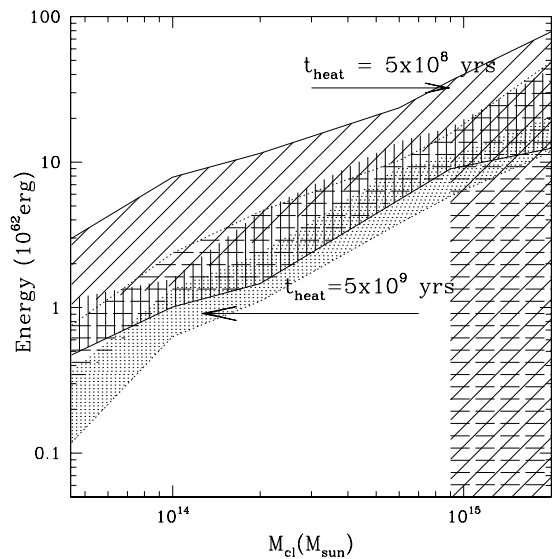}
\caption{Allowed ranges of total injected energy $E_{\rm agn}$ =
\lLr$\times\, t_{\rm \scriptscriptstyle heat}$ required to match
the observed entropy at $0.1r_{\rm \scriptscriptstyle 200}$ and/or
$r_{\rm \scriptscriptstyle 500}$, for two different values of
$t_{\rm \scriptscriptstyle heat}$, as a function of cluster mass.
The width of each region corresponds to the $1 - \sigma$ errors
plotted in Figure 1. The region bounded by thin solid lines
corresponds to the allowed range in energies for $t_{\rm
\scriptscriptstyle heat}=5\times 10^{8}$ years. Within this bounded
region, the area shaded with vertical solid lines is the spread in
energy that satisfies the entropy requirement at $0.1r_{\rm
\scriptscriptstyle 200}$ and the area shaded with oblique solid
lines is the spread in $E_{\rm agn}$ that satisfies the entropy
requirement at $r_{\rm \scriptscriptstyle 500}$. Only the overlap
region satisfies the observations at both radii. The region
bounded by thin dotted lines corresponds to the allowed range in
energies for $t_{\rm \scriptscriptstyle heat}=5\times 10^{9}$
years. In this region, the dotted area is the spread in energy
that satisfies the entropy requirement at $0.1r_{\rm
\scriptscriptstyle 200}$ and the area shaded with horizontal
long-dashed lines is the spread in $E_{\rm agn}$ that satisfies
the entropy requirement at $r_{\rm \scriptscriptstyle 500}$.}
\label{fig:Ej79_Mcl}
\end{figure}

\clearpage

\begin{figure}
\centering
\includegraphics[width=3.0in]{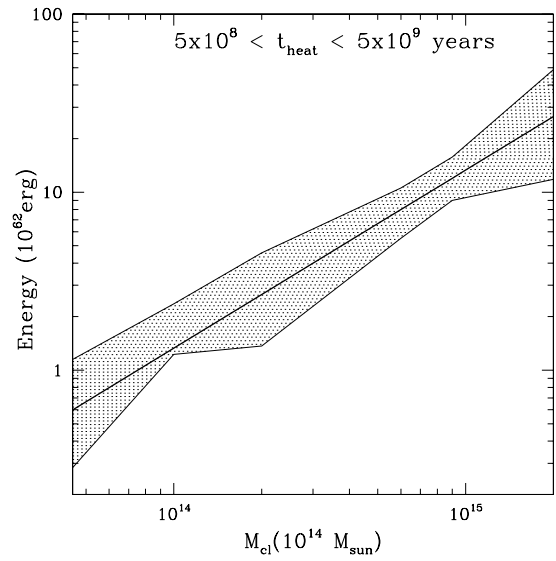}
\caption{This figure shows the permitted total injected energy range as a
function of the cluster mass for ICM heating times between $t_{\rm
heat}=5\times 10^{8}$yr (upper envelope) and $t_{\rm heat}=5\times 10^{9}$yr
(lower envelope). The shaded region  corresponds to values of $E_{\rm agn}$
that are able to match the  entropy  observations at {\it both} $0.1r_{\rm
\scriptscriptstyle 200}$  and $r_{\rm \scriptscriptstyle 500}$. The thick
solid line represents  a linear relation between the total energy injected
into the cluster by AGN and the mass of the cluster. The permitted
parameter space comes from the sum of permitted regions in Figure 
that satisfy the entropy constraints at both radii for fixed $t_{\rm heat}$.}
\end{figure}


\begin{thebibliography}{}
   \bibitem[Afshordi \& Cen(2002)]{afc02} Afshordi, N., \& Cen, R. 2002,
   \apj, 564, 669

   \bibitem[Allen \etal(2001)]{allen01} Allen, S.W. \etal 2001, \mnras,
   324, 842

   \bibitem[Babul \etal (2003)]{bblp02} Babul, A., Balogh, M.L., Lewis, G.F., \&
   Poole, G.B. 2002, \mnras, 330, 329

   \bibitem[Balogh \etal (1999)]{bbp99} Balogh, M.L., Babul, A., \& Patton, D.R. 1999, \mnras, 307, 463

   \bibitem[Basson \& Alexander (2003)]{ba03} Basson, J.F. \& Alexander, P. 2003,
   \mnras, 339, 353

    \bibitem[Binney \& Tabor
    (1993)]{bt95} Binney, J. \& Tabor, G. 1995, \mnras, 276, 663

   \bibitem[Begelman(2001)]{beg01} Begelman, M.C. 2001, in Gas and Galaxy
   Evolution, APS Conf. Proc., vol. 240, ed. Hibbard, J. E., Rupen,
   M.P., van Gorkom, J.H., p. 363, (astro-ph/0207656)

   \bibitem[Blanton \etal(2001)]{bsmcw01} Blanton, E.L., Sarazin, C.L.,
   McNamara, B.R., \& Wise, M.W. 2001, \apj, 558, 15

   \bibitem[Blanton \etal(2003)]{bsmc03} Blanton, E.L., Sarazin, C.L., \&
   McNamara, B.R. 2003, \apj, 585, 227

   \bibitem[Br\"uggen(2003a)]{br03a} Br\"uggen, M. 2003a, \apj, 593, 700

   \bibitem[Br\"uggen(2003b)]{br03b} Br\"uggen, M. 2003b, \apj, 592, 839

   \bibitem[Bryan(2000)]{bry00} Bryan, G.L. 2000, \apj, 544, 1

   \bibitem[Bullock \etal(2001)] {bull01} Bullock, J.S. \etal 2001,
   \mnras, 321, 559

   \bibitem[Churazov \etal(2001)]{ch01} Churazov, E. \etal 2001, \apj,
   554, 261

   \bibitem[Clarke \etal(2001)]{cnf94} Clarke, D.A. , Norman, M.L., \& Fiedler, R.A. 1994,
   ZEUS-3D User's Manual Version 3.2.1, (Urbana-Champaign: Univ. Illinois)


   \bibitem[Dav\'e \etal(2002)]{dkw02} Dav\'e, R., Katz, N., \& Weinberg, D.H.
   2002, \apj, 579, 23

   \bibitem[Ettori (2003)]{ett03} Ettori, S. 2003, \mnras, 344, 13

   \bibitem[Fabian \etal (2000)]{f00} Fabian, A.C. \etal 2000, \mnras,
   318, 65

   \bibitem[Ferrarese \& Merritt (2000)]{fm00} Ferrarese, L., \& Merritt,
   D. 2000, 539, 9

   \bibitem[Finoguenov \etal (2001)]{f01} Finoguenov, A., Reiprich, T.H., \&
   B\"ohringer, H. 2001, \aap, 368, 749

   \bibitem[Gebhardt \etal (2000)]{g00} Gebhardt, K., et al. 2000, \apj, 543, 5

   \bibitem[H\"aring \& Rix (2004)]{hr04} H\"aring, N., Rix, H.-W. 2004, ApJ, 604, 89

   \bibitem[Heinz \etal (2002)]{hcrb02} Heinz, S., Choi Y., Reynolds, C.S., \&
   Begelman M.C. 2002, \apj, 569, 79

   \bibitem[Johnstone \etal (2002)]{j03} Johnstone, R.M., Allen, S.W.,
   Fabian, A.C., \& Sanders, J.S. 2002, \mnras, 336, 299

   \bibitem[Kaiser \& Binney (2003)]{kai03} Kaiser, C.R., \& Binney, J. 2003,
   \mnras, 338, 837

   \bibitem[Kitayama \& Suto]{ks96} Kitayama, T., \& Suto, Y. 1996, \apj, 469, 480

   \bibitem[Komatsu \etal(2002)]{k02} Komatsu, E., \& Seljak, U. 2002,
   \mnras, 336, 1256

   \bibitem[Loken \etal(2000)]{ll00} Lloyd-Davies, E.J, Ponman, T.J.,
   Cannon, D.B. 2000, \mnras, 315, 689

   \bibitem[Loken \etal(2002)]{l02} Loken, C., Norman, M.L., Nelson, E.,
   Bryan, G.L., \& Motl, P. 2002, \apj, 579, 571

   \bibitem[Mazzotta \etal(2002)]{mz02} Mazzotta, P. \etal 2002, \apj,
   567, 37

   \bibitem[McNamara \etal(2000)]{mc00} McNamara, B.R. \etal 2000, \apj,
   534, 135

   \bibitem[McNamara \etal(2001)]{mc01} McNamara, B.R. \etal 2001, \apj,
   562, 149

   \bibitem[Muanwong \etal(2002)]{mtkp02} Muanwong, O., Thomas, P.A., Kay,
   S.T., \& Pearce F.R. 2002, \mnras, 336, 527

   \bibitem[Mushotzky \etal(2003)]{mu03} Mushotzky, R. \etal 2003,
   American Astronomical Society, HEAD meeting, 35, 13.07, astro-ph/0302267

   \bibitem[Nath(2003)]{nath03} Nath, B.B. 2003, \mnras, 339, 729

   \bibitem[Nath \& Roychowdhury(2002)]{nr02} Nath, B.B., \& Roychowdhury, S.
   2002, \mnras, 333, 145

   \bibitem[Omma et al. (2003a)]{o03a} Omma et al. 2003, \mnras, in press, astro-ph/0307471

   \bibitem[Omma \& Binney (2003b)]{o03b} Omma \& Binney, J., 2004, \mnras, astro-ph/0312658

   \bibitem[Ponman \etal(2003)]{pon03} Ponman, T.J., Sanderson, A.J.R.,
   \& Finoguenov, A. 2003, \mnras, 343, 331

   \bibitem[Peebles(1980)]{p80} Peebles, P.J.E. 1980, The Large Scale
   Structure of the Universe. Princeton Univ. Press, Princeton, NJ

   \bibitem[Peterson \etal(2001)]{p01} Peterson, J.R., \etal 2001, \aap,
   365, 104

   \bibitem[Quilis \etal(2001)]{qbb01} Quilis, V., Bower, R.G., \& Balogh, M.L.
   2001, \mnras, 328, 1091

   \bibitem[Reynolds \etal(2002)]{rhb02} Reynolds, C.S., Heinz, S., \&
   Begelman, M.C. 2002, \mnras, 332, 271

   \bibitem[Roychowdhury \& Nath(2003)]{rn03} Roychowdhury, S., \& Nath, B.B.
   2003, \mnras, 346, 199

   \bibitem[Ruszkowski \& Begelman(2002)]{rb02} Ruszkowski, M., \&
   Begelman, M.C. 2002, \apj, 581, 223

   \bibitem[Ruszkowski \etal(2004)]{rbb04} Ruszkowski, M., Br\"uggen, M.,
   \& Begelman, M.C. 2004, \apj, in press, astro-ph/0310760

   \bibitem[Stone \etal(1999)]{spb99} Stone, J.M., Pringle, J.E., \& Begelman,
   M.C. 1999, \mnras, 310, 1002

   \bibitem[Sutheland \& Dopita(1993)]{sud93} Sutherland, R.S., \&
   Dopita, M.A. 1993, \apjs, 88, 253

    \bibitem[Tabor \& Binney
    (1993)]{tb93} Tabor, G., \& Binney, J. 1993, \mnras, 263, 323

   \bibitem[Tornatore \etal
   (2003)]{tbsmmm03} Tornatore, L., Borgani, S., Springel, V., Matteucci, F.,
   Menci, N., \& Murante, G. 2003, \mnras, 342, 1025

   \bibitem[Tozzi \& Norman
   (2001)]{tn01} Tozzi, P., \& Norman, C. 2001, \apj, 546, 63

   \bibitem[Valageas \& Silk(1999)]{vs99} Valageas, P., \& Silk, J. 1999, \aap,
   350, 725

   \bibitem[Voit \& Bryan
   (2001)]{vb01} Voit, G.M., \& Bryan, G.L. 2001, \nat, 414, 425

    \bibitem[Wu \etal
    (2000)]{wfn00} Wu, K.K.S., Nulsen, P.E.J., \& Fabian, A.C.  2000,
    \mnras, 318, 889

    \bibitem[Wu \& Xue
    (2002)]{wx02} Wu, X., \& Xue, Y. 2002, \apj, 572, 19

\end{thebibliography}
\end{document}